\documentclass[trackchanges,twocolumn]{aastex701}
\usepackage[utf8]{inputenc}
\usepackage{textgreek}
\usepackage{amsmath}
\usepackage{amssymb}
\usepackage{pgfplots}
\usepackage{hyperref}
\usepackage{adjustbox}
\usepackage{tikz} 

\usetikzlibrary{positioning,calc}
\usepackage{pgfplots}
\usepackage{subcaption,graphicx}
\newsavebox{\tempfig}

\pgfplotsset{width=\paperwidth}
\pgfplotsset{compat=1.18}

\usepackage{xcolor} 
\usepackage[utf8]{inputenc}
\hypersetup{
    colorlinks=true, 
    linkcolor=blue, 
    citecolor=blue, 
    urlcolor=blue 
}

\newcommand{\ML}[1]{\textcolor{black}{#1}}

\begin{document}

\title{Machine Learning $\beta$-decay Half-lives and Their Application to $r$-Process Observables}

\author[0000-0002-6373-7494]{Mengke Li}
\email{mengkel@berkeley.edu}
\affiliation{Department of Physics, University of California, Berkeley, CA 94720, USA}
\affiliation{Department of Physics and Astronomy, University of Notre Dame, Notre Dame, IN, 46656, USA}

\author[0009-0008-6364-4547]{Flora Wang}
\email{florayw@berkeley.edu}
\affiliation{Department of Astronomy, University of California, Berkeley, CA 94720, USA}

\author[0000-0002-2748-6640]{Jonathan Engel}
\email{engelj@physics.unc.edu}
\affiliation{Department of Physics and Astronomy, University of North Carolina, Chapel Hill, NC 27516, USA}

\author[0000-0002-9950-9688]{Matthew Mumpower}
\email{Matthew Mumpower matthew@mumpower.net}
\affiliation{Obsidian Research, Fort Wayne, IN 46835, USA}
\affiliation{Department of Physics and Astronomy, University of Notre Dame, Notre Dame, IN, 46656, USA}

\author[0000-0002-4729-8823]{Rebecca Surman}
\email{Rebecca Surman rsurman@nd.edu}
\affiliation{Department of Physics and Astronomy, University of Notre Dame, Notre Dame, IN, 46656, USA}

\author[0000-0002-3305-4326]{Nicole Vassh}
\email{Nicole Vassh nvassh@gmail.com}
\affiliation{TRIUMF, 4004 Wesbrook Mall, Vancouver, British Columbia V6T 2A3, Canada}

\date{\today}

\begin{abstract}
Accurately modeling $\beta$-decay half-lives of highly neutron-rich nuclei is a critical challenge for understanding $r$-process nucleosynthesis and the resulting kilonova light curves. We introduce a data-driven approach to model $\beta$ decay that uses Mixture Density Networks (MDNs) trained on the latest experimental measurements to extrapolate the half-lives of neutron rich nuclei. Unlike standard deterministic regression, the MDN directly parameterizes the probability distribution of the half-lives, providing intrinsic (aleatoric) uncertainty. We contrast the intrinsic uncertainty of a single model with the range of models produced by using random variations of the input training data samples. While all models show excellent agreement with experimental data, we find that varying the training data sets produces a wide range of extrapolations, particularly along data-poor regions such as the $N=126$ isotonic chain.   We then propagate a selection of model results into $r$-process network calculations to evaluate their impact on isotopic abundances. Finally, by coupling the resulting isotopic abundances with thermalization efficiencies, we translate the effective nuclear heating rates into bolometric light curves. We compare the ranges of outcomes produced from the intrinsic uncertainty of a single model to those produced by 
multiple independent models trained on different training data sets.
\end{abstract}


\section{Introduction}

The origin of the heavy elements is a long-standing open question in physics.
While the rapid neutron capture process ($r$-process) is understood to be responsible for synthesizing approximately half of the elements heavier than iron, the precise astrophysical site at which the nucleosynthesis takes place and the detailed path it follows remain subjects of debate \citep{1957RvMP...29..547B, 1965ApJS...11..121S}. The detection of the binary neutron-star merger GW170817 provided the first direct evidence of $r$-process nucleosynthesis in situ \citep{Abbott2017, Cote_2018}; however, decoding the details of such events requires a precise mapping of the nuclear properties of isotopes far from stability. Because these neutron-rich nuclei lie in the region of the nuclear chart where experimental data is scarce, astrophysical predictions must rely heavily on theoretical input \citep{Mumpower_2016}.

Among the requisite nuclear properties, the $\beta$-decay half-life is critical. 
As the process responsible for converting neutrons into protons, $\beta$ decay drives the nuclear flow toward higher atomic numbers. Its role becomes particularly important at the nuclear shell closures ($N=82$ and $N=126$), where the ``waiting point" approximation holds: the $r$-process flow stalls as nuclei wait to decay, leading to the buildup of material that forms the characteristic abundance peaks. 
As the temperature drops, once the equilibrium between neutron captures and photodissociations fails, $\beta$ decay competes with the last neutron captures to finalize the abundance pattern. Beyond shaping nucleosynthesis yields, $\beta$-decay plays a vital role in the era of multi-messenger astronomy: it is the primary source of radioactive heating that powers the electromagnetic transient of the kilonova, except in cases with significant late-time fission \citep{Zhu2018, Zhu_2021}. Consequently, uncertainties in $\beta$-decay rates propagate directly into uncertainties in the kilonova light curve, affecting our ability to interpret observations from current and future telescopes \citep{Lund_2023}.

Accurately calculating $\beta$-decay rates across the entire nuclear chart remains a theoretical challenge. Such calculations demand a precise description of the nuclear ground state, excited states in the daughter nucleus, and the transitions connecting them. The pioneering work of \citet{Moller2003} (hereafter MLR), established a framework for global predictions based on the quasiparticle random phase approximation (QRPA). More recently, the field has been advanced by fully self-consistent approaches, including covariant density functional theory (CDFT) combined with QRPA, developed by \citet{Marketin_2016} (hereafter MKT), Skyrme energy-density functionals with the Finite Amplitude Method (FAM) by \citet{Ney_2020} (hereafter Ney), and recent large-scale microscopic calculations by \citet{ravlic_2025}, which yield results similar to Ney. 
These calculations provide invaluable insights, and we will compare our approach with MLR, MKT, and Ney in the following sections.

As a complementary method for prediction across the nuclear chart, Machine Learning (ML) can be a powerful tool, capable of capturing complex, non-linear dependencies in the data.
Previous works have successfully introduced data-driven approaches to make half-life predictions \citep{You_2025, Jyothish_2025, Yuan_2026,Sihem_2026, Jain_2026}. For $\beta^-$ decay modeling specifically, \citet{Niu_2019} used Bayesian Neural Networks (BNNs) to predict half-lives with quantified uncertainties for $r$-process abundance calculations; \citet{Amir_2025} employed Support Vector Machines (SVMs) to improve predictive accuracy; and \citet{Li_ML_decay_2026} incorporated physics-informed machine learning to predict half-lives and evaluate them against newly measured experimental data.
To complement existing ML approaches, we adapt the Mixture Density Network (MDN) \citep{MDN} method, previously used for nuclear mass modeling \citep{mumpower_2023,LI_2024}, to evaluate $\beta$-decay half-lives. 
By directly learning the probability distribution of the half-life, the MDN provides a natural, data-driven way to quantify the evolution of predictive uncertainty as one extrapolates from long-lived isotopes toward the neutron drip line. We will use the half-life probability distributions in assigning uncertainty to calculate heavy-element abundances and electromagnetic observables.

The remainder of this paper is organized as follows: Section \ref{sec:methods} outlines the data-processing pipeline for MDN training and describes the way we simulate nuclesynthesis.  We then present our results in three distinct subsections of Section \ref{sec:Results}. First, Section \ref{subsec:model_perform} evaluates the overall model performance, demonstrating that our MDN framework accurately predicts known half-lives while generating physically bounded uncertainty. Next, Section \ref{subsec:r-process} integrates these rates into $r$-process network calculations to evaluate their impact on isotopic abundances under both merger dynamical ejecta and disk wind conditions. Finally, Section \ref{subsec:light_curves} propagates the nuclear uncertainties to simulate their effect on the bolometric luminosity of the resulting kilonova, providing an uncertainty-quantified estimate of the light curve band arising from fundamental unknowns in exotic $\beta$-decays.

\section{Methods}\label{sec:methods}

\subsection{Input Features}

Good performance by an MDN model relies critically on the selection of an informative feature space.
To construct a physics-informed feature space for predicting $\beta$-decay half-lives ($\log_{10}(T_{1/2})$), we start by choosing the proton number ($Z$) and the neutron number ($N$) as baseline features.
Next, we include the decay energy ($Q_\beta$), sometimes called the $Q$-value.
In the Fermi theory of $\beta$-decay, the decay rate ($\lambda = \ln 2 / T_{1/2}$) is governed largely by the available leptonic phase space, which scales approximately with the fifth power of the decay energy ($Q_\beta^5$). Consequently, $Q$-values are an essential input feature for the neural network to properly learn about decay trends.

Standard global models often struggle to reproduce the odd-even staggering in half-lives, which arises from nucleon-pairing interactions. To explicitly inform the model about staggering, we introduce proton ($P_Z$) and neutron ($P_N$) pairing terms. 
Because the impact of pairing correlations tends to diminish with increasing mass, we scale these terms inversely with the nucleon number:
\begin{align}
    P_Z &= \begin{cases} 1/Z & \text{if } Z \text{ is even} \\ -1/Z & \text{if } Z \text{ is odd} \end{cases} \\
    P_N &= \begin{cases} 1/N & \text{if } N \text{ is even} \\ -1/N & \text{if } N \text{ is odd} \end{cases}
\end{align}
This formulation captures odd-even staggering while naturally dampening its magnitude for heavy nuclei.

To capture the sharp changes in nuclear structure near magic numbers, we include the one-neutron pairing metric, $D_n$:
\begin{equation}
    D_n(Z, N) = (-1)^{N-1} \left( S_n(Z, N+1) - S_n(Z, N) \right)
\end{equation}
This quantity can serve as a probe of shell closures because the magnitude of $D_n$ exhibits a distinct spike when the neutron number is at a closed shell \citep{Vassh_2021}, effectively locating the ``waiting points" in the nuclear chart.
The complete input feature space is summarized in Table~\ref{tab:features}.

\begin{table}[h]
    \setlength{\tabcolsep}{3pt}
    \centering
    \caption{Summary of input features used for the MDN model.}
    \label{tab:features}
    \begin{tabular}{l|l}
    \hline
    \textbf{Feature} & \textbf{Description} \\
    \hline
    $Z$ & Proton number \\
    $N$ & Neutron number \\
    $Q_\beta$ & Beta-decay energy ($M(Z+1,N-1) - M(Z, N)$) \\
    $P_Z$ & Proton pairing term (scales with $1/Z$) \\
    $P_N$ & Neutron pairing term (scales with $1/N$) \\
    $D_n$ & Separation energy difference (Shell indicator) \\
    \hline
    \end{tabular}
\end{table}

\subsection{Data Preparation}

We construct the primary dataset for experimental $\beta$-decay half-lives ($T_{1/2}$) from the NUBASE2020 evaluation \citep{Kondev_2021_Nubase2020}. 
To focus the study strictly on $\beta^-$ unstable nuclei, we evaluate the branching ratios; for nuclei with competing decay channels like $\alpha$ emission, we adjust only the half-life contribution associated with the $\beta^-$ decay branching.
We then apply two primary selection criteria: we restrict the sample to isotopes with half-lives $T_{1/2} < 10^6$ s, and we include only nuclei with proton and neutron numbers $Z, N \ge 8$, so as to exclude very light nuclei. 
These filtering steps yield a final dataset comprising 1279 nuclei.

We calculate input quantities ($Q_\beta$ and $D_n$) primarily from experimental atomic masses in the AME2020 database \citep{Wang_2021_AME}.  Because experimental mass measurements do not fully cover all known half-lives, however, some neutron-rich nuclei in NUBASE2020 lack experimental masses. 
For these nuclei, we calculate $Q_\beta$ and $D_n$ by using theoretical mass predictions from three global models: the Finite-Range Droplet Model (FRDM) \citep{FRDM2012}, the Hartree-Fock-Bogoliubov model (HFB) \citep{Goriely_HFB}, and the Weizsäcker-Skyrme model (WS4) \citep{WANG_WS4}.

To incorporate experimental uncertainties into training, we implement a Monte Carlo data augmentation strategy.  
Instead of treating inputs and targets as fixed point values, we treat them as Gaussian distributions, defined by their experimental mean values and uncertainties. 
Specifically, we generate $18$ augmented samples for every nucleus. We also sample the target half-lives ($\log_{10}( T_{1/2})$) and the input masses (used to calculate $Q_\beta$ and $D_n$) from Gaussian distributions centered at their experimental values.
This sampling forces the network to learn that data points with large experimental errors are less constraining than precise measurements. 

For nuclei whose masses rely on theoretical models, we assign a systematic mass uncertainty of 0.1 MeV. In such cases, we split the 18 samples evenly among the theoretical models (six samples each from FRDM, HFB, and WS4), ensuring balanced representation.

\begin{figure*}[ht!]
\centering
\includegraphics[width=0.95\linewidth]{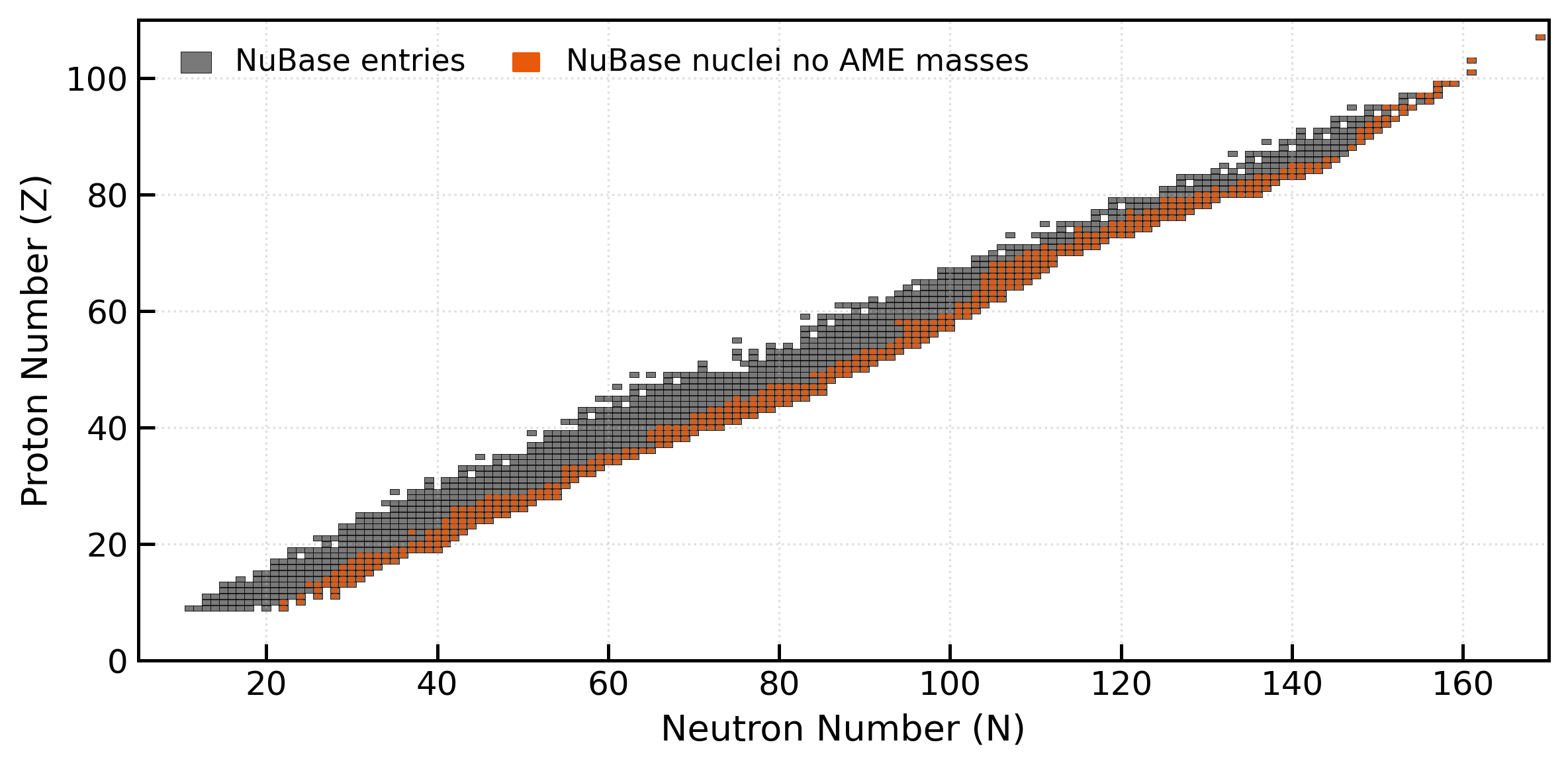}
\caption{
Representation in the nuclear chart of the NUBASE2020 dataset used in this work. Gray markers denote nuclei with available experimental masses (AME2020), while orange markers indicate neutron-rich nuclei where theoretical mass models (FRDM, HFB, WS4) were used to compute input features.
}
\label{fig:nubase}   
\end{figure*}

The resulting dataset is represented visually in Fig.\ \ref{fig:nubase}. 
The complete NUBASE2020 dataset is highlighted in gray. 
Orange squares represent nuclei for which AME masses are unavailable. 
Those are primarily situated along the neutron-rich boundary,
where theoretical mass models fill the data gaps.

\subsection{Mixture Density Network Implementation}

We employ a Mixture Density Network (MDN) to map the input features to the target half-lives. Unlike standard regression models that produce a single deterministic value, an MDN outputs the parameters of a probability distribution, specifically, the means ($\mu$), standard deviations ($\sigma$), and mixing coefficients ($\pi$) of a Gaussian mixture. This probabilistic framework allows us to model data distributions and, crucially, provides a quantification of the predictive uncertainty intrinsic to the model.

\begin{figure*}[t]
    \centering
    \includegraphics[width=0.95\linewidth]{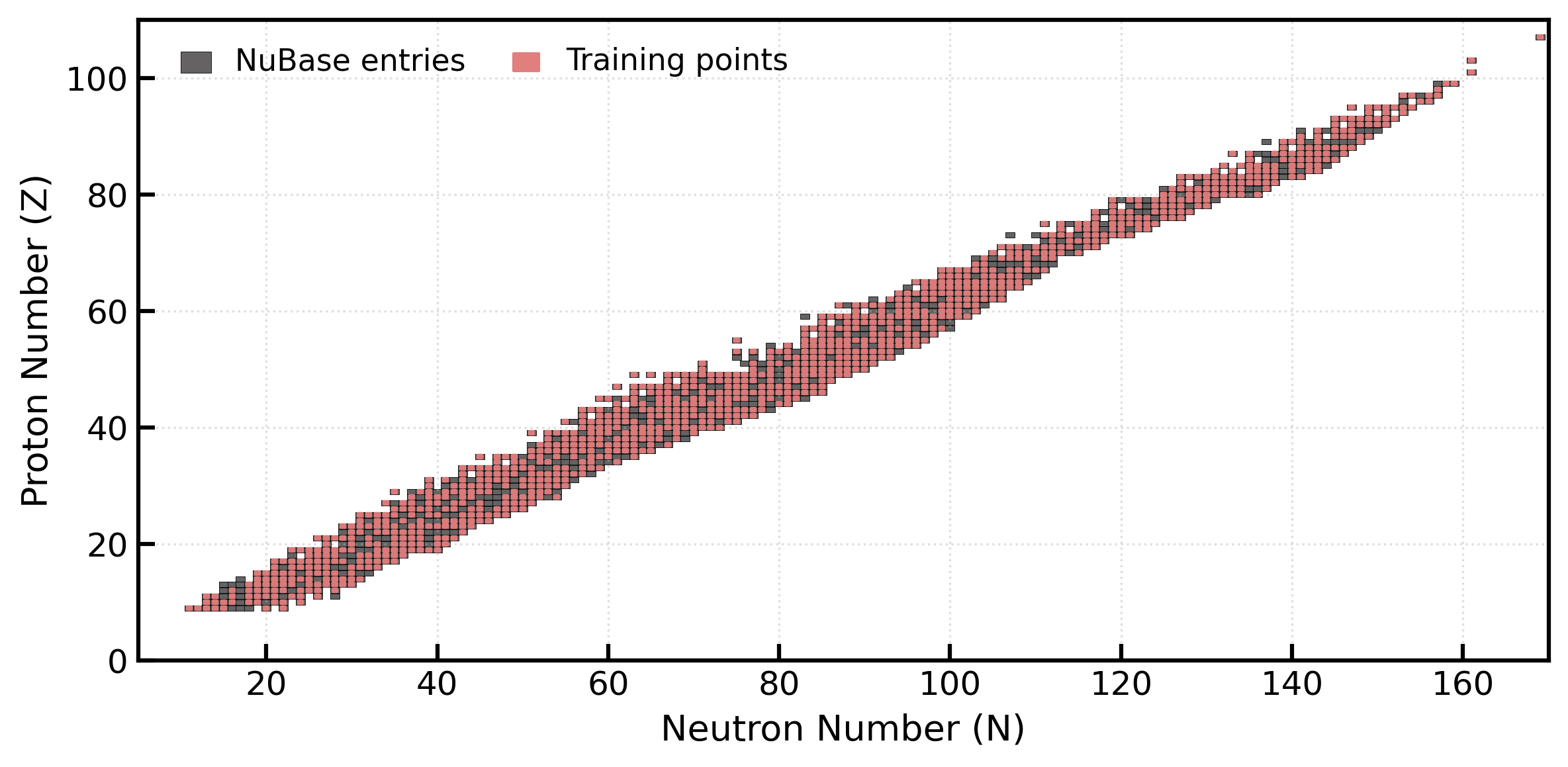}
    \caption{Distribution of the training and testing sets. The pink squares represent the training nuclei, while the gray squares represent the testing set used for validation.}
    \label{fig:train:test}   
\end{figure*}

We use a random stratified split to divide the data, designating 80\% of it for training and the remaining 20\% for testing. The distribution of these subsets is depicted in Fig.\ \ref{fig:train:test}, where training points are highlighted in pink and testing points in gray.

The neural network architecture consists of an input layer accepting the six features listed in Table \ref{tab:features}, followed by four hidden layers with 32 neurons each. We use the hyperbolic-tangent (tanh) activation function to handle the non-linearities in the nuclear landscape, and use the Adam algorithm \citep{Adam_optim} to optimize the model.

The objective function is the negative log-likelihood of the Gaussian mixture, supplemented by a regularization term to prevent the network from predicting physically unrealistic (infinite) uncertainties. The general loss function is defined as:
\begin{equation}
\begin{split}
    \mathcal{L} = & - \sum_{i} \log \left( \sum_{k} \pi_k(x_i) \mathcal{N}(y_i | \mu_k(x_i), \sigma_k(x_i)) \right) \\
    & + \lambda_{\text{reg}} \sum_k \sigma_k
\end{split}
\end{equation}
Here, we find that the distribution of half-lives is sufficiently unimodal to allow a single Gaussian kernel ($k=1$) to capture the data structure effectively. Consequently, we set the mixing coefficient $\pi_k$ to 1, as in previous work \citep{Lovell_2022, Mumpower_2022, mumpower_2023, LI_2024}, and the model focuses on learning the mean $\mu$ and the uncertainty $\sigma$. 

To balance computational efficiency with model stability, we set the variance regularization weight to $\lambda_{\text{reg}} = 0.1$. Although generating 18 samples per $(Z, N)$ pair sufficiently reduces training time compared to e.g., 100 samples, it makes the unregularized model predict very large uncertainties for certain isotopes. The regularization term mitigates these sparse-sampling artifacts.
This weighting is optimal for the training dynamics: it contributes $\sim$6\% to the total loss initially, allowing the negative log-likelihood to dominate early learning, and subsequently increases in relative importance to provide crucial stabilization during fine-tuning.

\subsection{Astrophysical Simulation Framework}

To evaluate the impact of the MDN-predicted half-lives on physical r-process observables, we use the PRISM nuclear reaction network \citep{Sprouse_2021} to perform $r$-process nucleosynthesis calculations. For the nuclear physics inputs, we employ a consistent baseline to isolate the effects of the $\beta$-decay rates. We take masses and neutron capture rates from FRDM and CoH calculations \citep{FRDM2012, Kawano2016_CoH}, respectively, and model fission fragment distributions and fission rates following \citet{Vassh_2019}. The thermodynamic evolutions come from neutron-star merger simulations for dynamical ejecta \citep{Ross_2014} and accretion disk ejecta \citep{Lund_2024}. 

Following the nucleosynthesis simulations, we determine the evolution of the effective nuclear heating by connecting the energy released from radioactive decays (the $Q$-values) to the efficiency with which the decay products thermalize in the expanding ejecta. 
With the methodology outlined in \citet{Lund_2023, KB_2019, Zhu_2021}, the total effective heating rate, $\dot{Q}(t)$, is given by:
\begin{equation}
    \dot{Q}(t) = \sum_{i} \dot{q}_i(t) f_i(M_{\text{ej}}, v_{\text{ej}}, t) M_{\text{ej}}
\label{eq:Q_dot}
\end{equation}
where the sum covers all active decay channels. Here, $\dot{q}_i(t)$ is the specific energy generation rate, and $f_i(M_{\text{ej}}, v_{\text{ej}}, t)$ is the time-dependent thermalization efficiency. Following \citet{KB_2019}, we take these efficiencies to depend on macroscopic outflow properties; in our model, we use an ejecta mass $M_{\text{ej}}$ = 0.05 $M_\odot$ and expansion velocity $v_{\text{ej}}$ = 0.15$c$.

To translate these effective heating rates into bolometric light curves, we model the ejecta as a homologously expanding envelope in the semi-analytic framework of \citet{KB_2019} and \citet{Lund_2023}. To account for the high opacity of the lanthanide-rich composition in the ejecta, we treat opacity as temperature dependent, with a maximum value of $\kappa_{\max} = 100$ cm$^2$ g$^{-1}$, as in the treatment of \citet{Lund_2023}. 

\section{Results}\label{sec:Results}
In this section, we present the results of our MDN computations, evaluating their performance in predicting $\beta$-decay half-lives and propagating predictions of select models into calculations of $r$-process nucleosynthesis and the resulting kilonova light curves. 

\subsection{Model performance}
\label{subsec:model_perform}

Here we select a single model and evaluate the accuracy of its predictions in the regions of the nuclear chart where information is known.
We compare the selected MDN's performance on the training set (80\% of NUBASE2020 data) with its performance on the independent testing set (20\%). 
The model achieves a training root-mean-squared error (RMSE) of 0.604 and a testing RMSE of 0.566, resulting in an overall RMSE of 0.597 for the entire NUBASE2020 dataset. 
The close agreement between the training and testing metrics indicates that the model has not over-fitted and that our training/testing sets separately represent the overall data well.

We next compare our global predictions for beta-decay half-lives from machine learning to the predictions of several theoretical models. For context, we note that the MLR \citep{Moller2003}, MKT \citep{Marketin_2016}, and Ney \citep{Ney_2020} models achieve RMSE values of approximately 0.82, 1.39, and 0.82, respectively. The reduction in RMSE by roughly 30\% relative to the theoretical models demonstrates that the framework properly incorporates known trends in beta-decay half-lives.

\begin{figure}[h!]
\centering
\includegraphics[width=\linewidth]{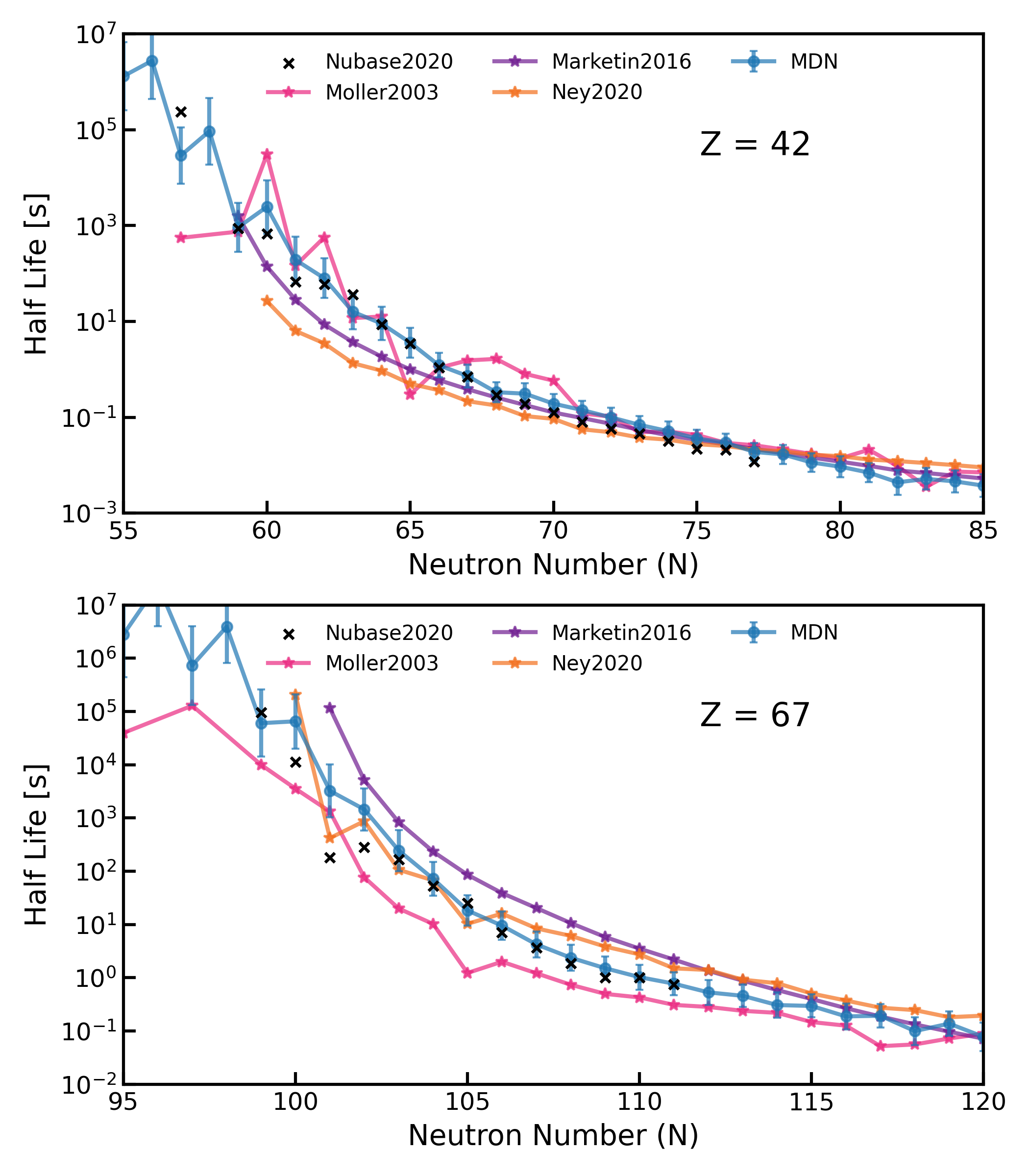}
\caption{
Half-life predictions for the $Z = 42$ (top) and $Z = 67$ (bottom) isotopic chains. The MDN (blue solid line) is compared with experimental data (black) and theoretical models: MLR (magenta), MKT (purple), and Ney (orange). 
}
\label{fig:hl_Z}   
\end{figure}

Although global metrics such as the RMSE demonstrate the average predictive accuracy of the MDN in nuclei with known rates, it is crucial to verify that the network correctly captures localized physical trends in addition to fitting the bulk data. 
To assess this local predictive capability, we analyze specific isotopic chains.
Fig.\ \ref{fig:hl_Z} displays the $\beta$-decay half-lives for the Molybdenum ($Z=42$) and Holmium ($Z=67$) isotopes. 
The MDN predictions (solid blue line) show excellent agreement with the experimental data points from NUBASE2020, accurately reproducing the overall trend toward shorter half-lives with increasing neutron number. 
The error bars represent the $1\sigma$ uncertainty derived directly from the MDN output variance ($\sigma$). 

\begin{figure}[h!]
\centering
\includegraphics[width=\linewidth]{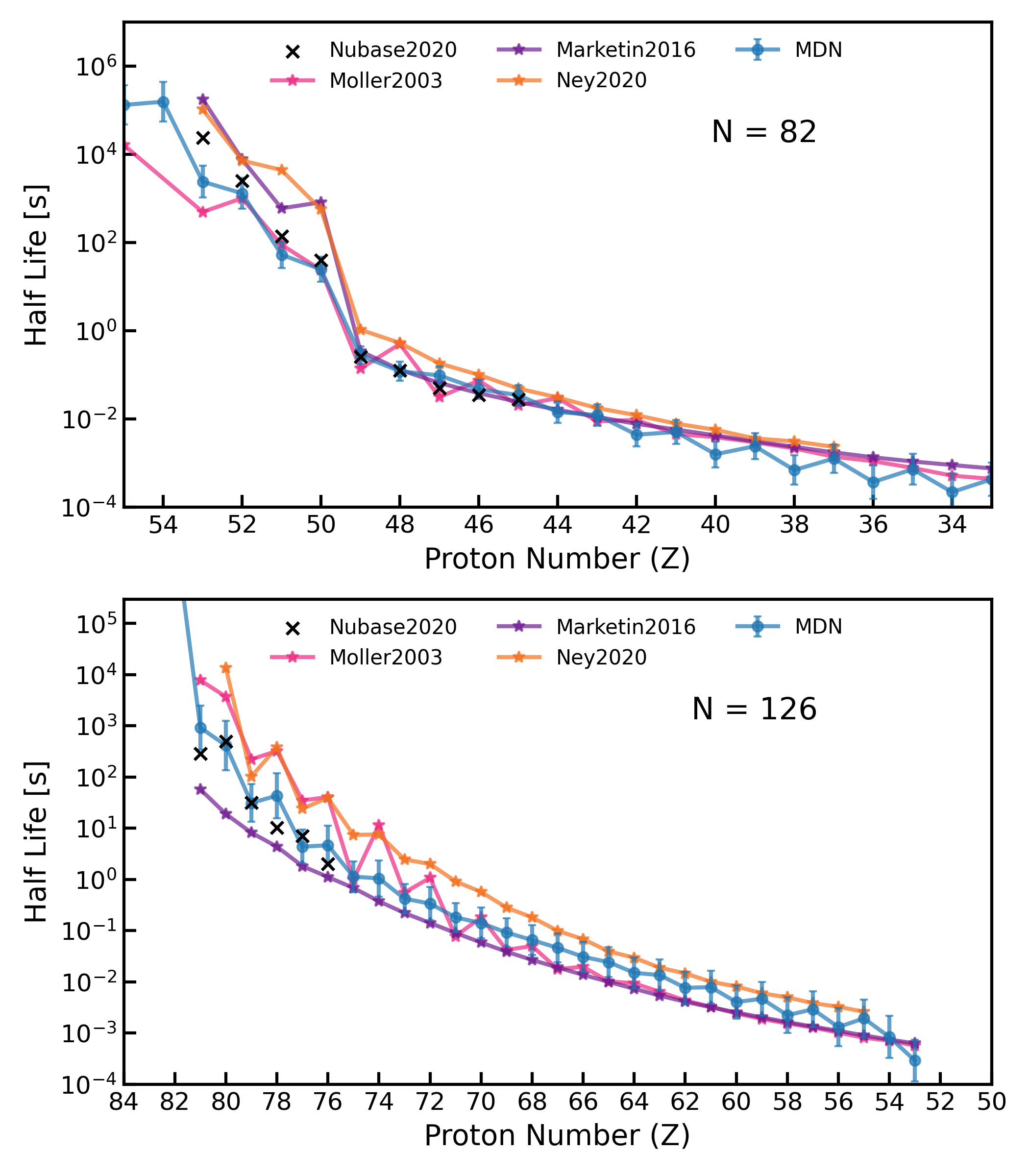}
\caption{
Half-life predictions for the $N=82$ (top) and $N=126$ (bottom) isotonic chains. The MDN (blue solid line) is compared with experimental data (black) and theoretical models: MLR (magenta), MKT (purple), and Ney (orange).
}
\label{fig:hl_N}   
\end{figure}

Examining whether isotopic trends are reproduced is not only important for model validation, but also for assessing predictions in the neutron-rich regions accessed by the $r$ process. Another critical diagnostic is model behavior along isotonic chains, in particular at neutron shell closures.
These ``magic-number" nuclei act as the primary waiting points in the $r$-process, where the nuclear flow stalls and material accumulates to form the characteristic abundance peaks \citep{Kuske_2025, LI_2026}.  Any systematic bias in predicted half-lives at these closures will directly affect the resulting abundance pattern.
Fig.\ \ref{fig:hl_N} presents the half-life predictions for the $N=82$ (top panel) and $N=126$ (bottom panel) isotonic chains.

In the $N=82$ region, because of the relatively high density of experimental data, the MDN predictions agree well with both the measurements and the predictions of theoretical models (MLR, MKT, Ney). However, an important difference emerges in the region beyond current data ($Z < 40$). Here, the MDN predicts persistent, strong odd-even staggering, whereas several theoretical models tend to smooth out these variations. 

Several targeted tests show that this behavior stems from the pairing input features ($P_Z, P_N$); removing them eliminates the staggering, while using unattenuated constant pairing terms causes very strong staggering along isotopic chains. By instead utilizing our suppressed pairing terms, the network produces the odd-even variations near known data while gently dampening them as the model extrapolates. The remaining staggering far from stability is still larger, however, than that produced by the theoretical models.

The $N=126$ shell closure presents a greater challenge because experimental constraints are more scarce. As shown in the bottom panel of Fig.\ \ref{fig:hl_N}, the MDN captures the experimental trend at this shell closure more accurately than the theoretical models, which overestimate or underestimate the magnitude of the half-lives.

Crucially, we observe a divergence in model predictions in the $N = 126$ isotonic chain as we move from higher $Z$ to lower $Z$. At higher proton numbers ($66 < Z < 75$), the MDN aligns more closely with the shorter half life MLR/MKT predictions. In the lower-$Z$ region (representing very neutron-rich nuclei), however, the MDN predictions diverge from the MLR/MKT predictions and tend to approach those of the Ney model. 
This shift indicates a strong $Z$-dependence in the MDN's extrapolation, resulting in systematically longer half-lives in lower Z, neutron-rich nuclei. 
As a result, the pattern of half lives at the $126$ shell closure in the MDN model is different from the smooth extrapolation trends that characterize the other global models.

\begin{figure}[h!]
\centering
\includegraphics[width=\linewidth]{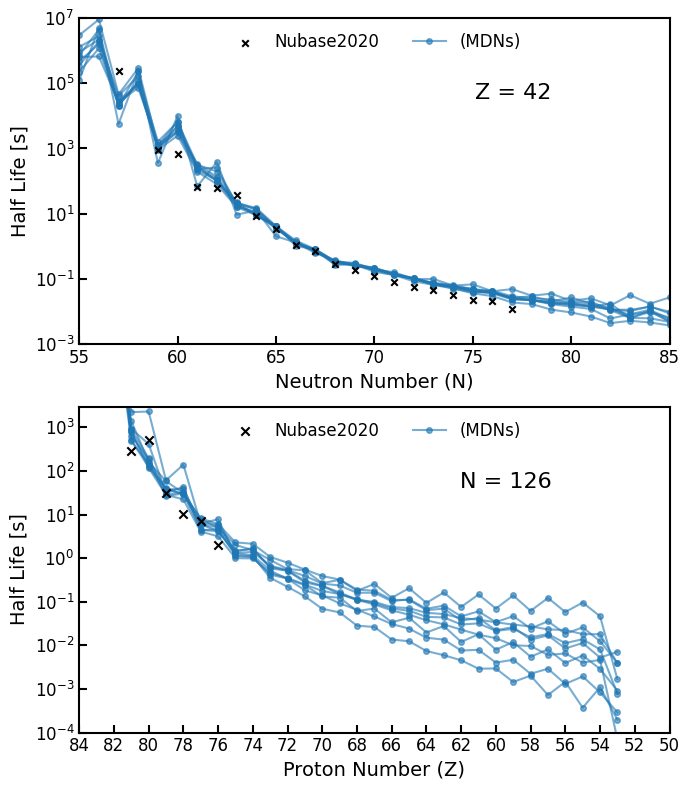}
\caption{
$\beta$-decay half-life predictions for the $Z = 42$ isotopic chain (top panel) and the $N = 126$ isotonic chain (bottom panel). The blue lines represent an ensemble of six MDN models, trained using different random seeds to generate distinct training data samples. Experimental data from NUBASE2020 are shown for reference.
}
\label{fig:diff_seeds}   
\end{figure}

Given the paucity of nuclear data near $N=126$, the extrapolation away from stability for this isotone depends strongly on the choice of input half lives. We test this dependence by selecting different random seeds at the start of the training, so that the training data set varies. The results appear in Fig.~\ref{fig:diff_seeds}. The random variations of training data don't strongly affect the network's $\log_{10}(T_{1/2})$ RMSE values, measured against the known NUBASE2020 \citep{Kondev_2021_Nubase2020} experimental data. Predictions in data-rich regions have the same stability; as the figure shows, the model ensemble exhibits only slight variations along the $Z=42$ isotopic chain. By contrast, predictions in data-poor, neutron-rich regions far from stability, highlighted by the $N=126$ isotonic chain in the bottom panel, demonstrate a strong dependence on the random seed, causing the extrapolations to diverge substantially. We examine the astrophysical impact of this variation in the next section.

\begin{figure}[h!]
\centering
\includegraphics[width=\linewidth]{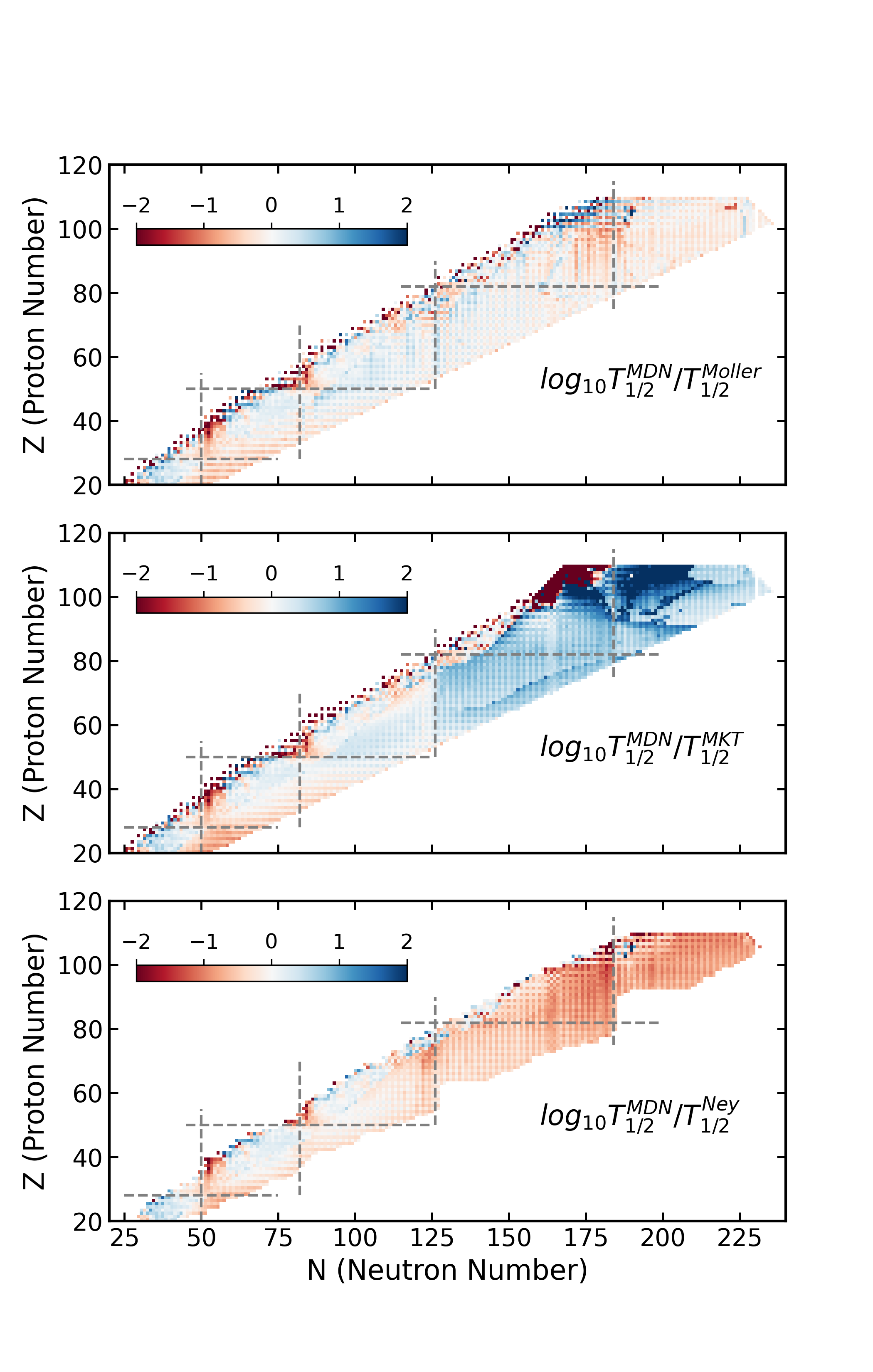}
\caption{
The predicted beta decay half life for the whole nuclear chart compared with those of MLR, MKT, and Ney. 
}
\label{fig:hl_whole_chart}   
\end{figure}

Having examined the local isotopic trends we return to our base MDN model and extend our analysis to the whole nuclear chart. Figure \ref{fig:hl_whole_chart} presents the ratio of the MDN-predicted half-lives to the experimental values from NUBASE2020 (where available) and compares our extrapolations against the theoretical models MLR, MKT, and Ney. 
On average, the MDN predictions align most closely with the MLR model across the chart, though with scattered differences, as shown in the top panel.

Distinct systematic trends emerge when comparing the MDN with theoretical models. In the lighter-mass region, the MDN generally favors shorter half-lives compared to the MKT model. However, this behavior inverts in the heavy nuclei, specifically beyond the $N=126$ shell closure, where the MDN predicts significantly longer half-lives than MKT, as the middle panel of Fig.\ \ref{fig:hl_whole_chart} shows. 
This long decay timescale has profound implications for nucleosynthesis, because faster decay in this region can cause fission to reduce actinide abundances early, directly influencing the final $r$-process abundance pattern.
Conversely, when compared to the microscopic Ney model, the MDN generally predicts shorter half-lives across the neutron-rich nuclei, effectively positioning our predictions between the two distinct theoretical predictions.

\begin{figure}[h!]
\centering
\includegraphics[width=\linewidth]{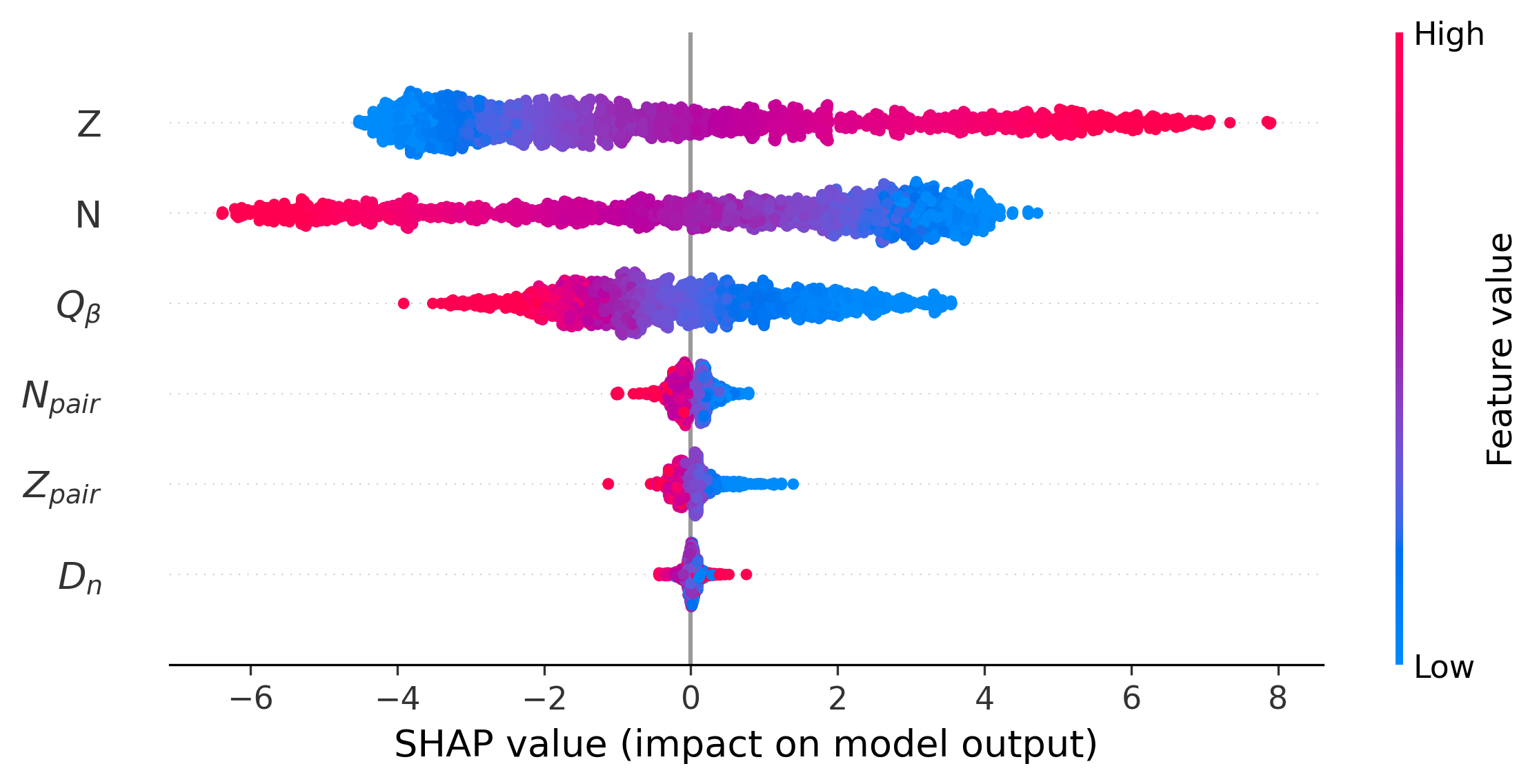}
\caption{
The feature importance as quantified by the SHAP values, detailing the influence of each feature in model predictions.
}
\label{fig:shap}   
\end{figure}

To understand the decision-making of our network, we use SHAP (SHapley Additive exPlanations) values to rank feature importance (Fig.\ \ref{fig:shap}) \citep{SHAP_2017, Mumpower_2022}. 
They reveal that the dominant physical features governing $\beta$-decay predictions are proton number $Z$, neutron number $N$ and the $\beta^-$ decay $Q_\beta$ values, which are the three most influential features. 
First, high $Z$ and low $N$ values have a positive contribution to the output, pushing the model toward longer half-lives, as in nuclei closer to the valley of stability.
Low $Z$ and high $N$ combinations, by contrast, correspond to a large neutron excess and negative SHAP values, leading to shorter half-lives.
Next, high $Q_\beta$ values (associated with unstable, neutron-rich nuclei) make the half-lives shorter. This behavior agrees qualitatively with the general trend of Sargent’s law ($1/T_{1/2} \propto Q_\beta^5$), even though the model does not enforce the power law.
The pairing energy terms are also significant, enabling the model to reproduce the odd-even staggering discussed in previous section. 
Finally, the neutron shell closure information ($D_n$) provides necessary corrections in the shell regions, modulating the bulk trends near closed shells.

\subsection{Astrophysical Application: $r$-Process Nucleosynthesis}
\label{subsec:r-process}

To quantify the impact of our machine-learning half-life predictions on heavy element synthesis, we integrate the MDN-predicted $\beta$-decay rates into theoretical $r$-process network calculations. 

We specifically examine cold dynamical ejecta contained in the tidal tails ejected during neutron star mergers.  We adopt the trajectory of \citet{Ross_2014}, characterized by a low entropy and low initial electron fraction ($Y_e \approx 0.02$). This scenario is particularly useful for isolating the effects of $\beta$-decay; the rapid expansion and low entropy cause photodissociation rates to drop off early, breaking $(n, \gamma) \rightleftarrows (\gamma, n)$ equilibrium \citep{LI_2022}. As a result, the final abundance pattern is shaped directly by the competition between neutron capture and $\beta$-decay, making it a sensitive probe of these underlying nuclear-physics inputs.

\begin{figure}[t]
    \centering
    \includegraphics[width=\linewidth]{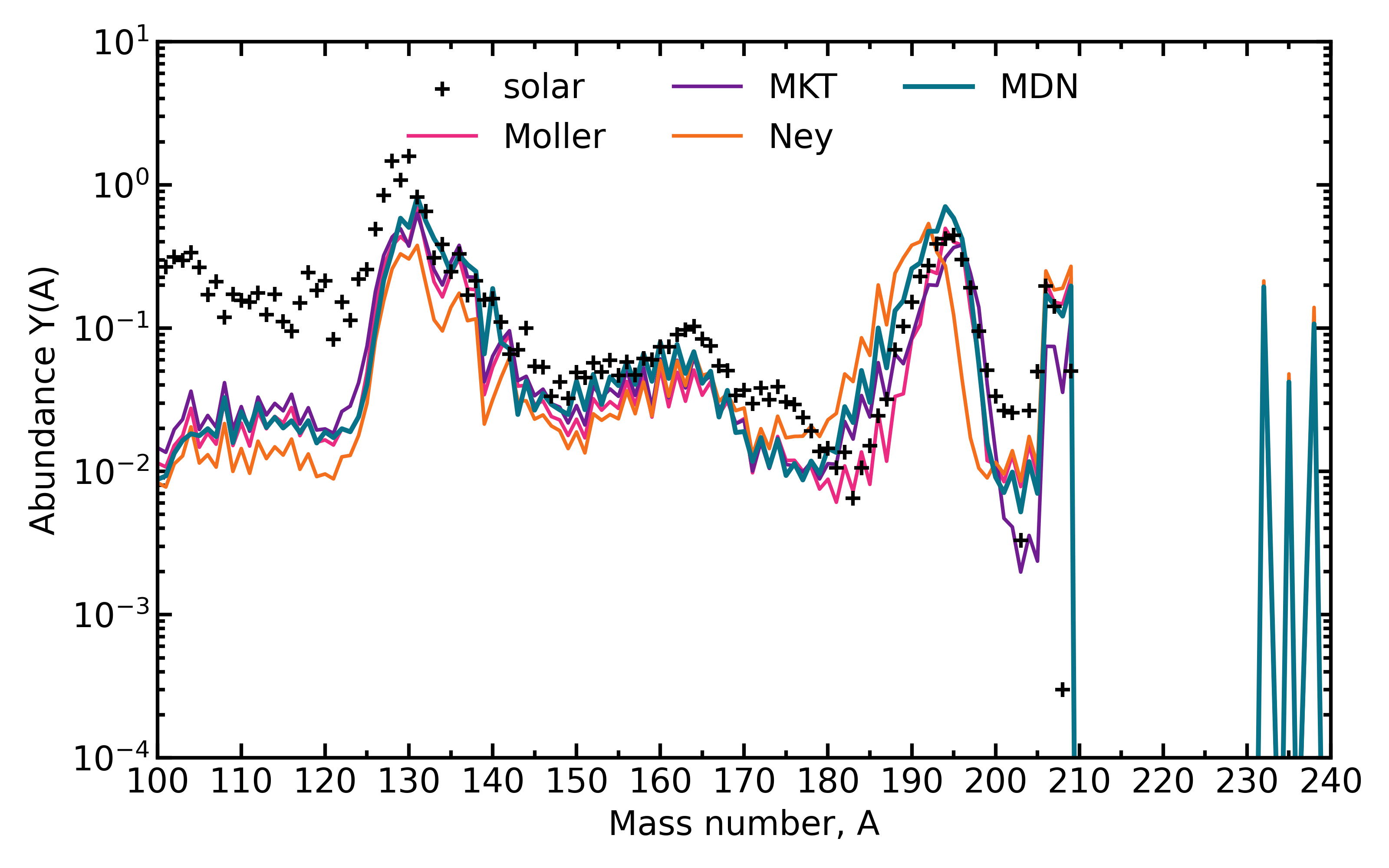}
    \caption{Simulated $r$-process abundance patterns with cold dynamical ejecta.  The teal line represents the results of the calculation with mean MDN rates, while the other lines are those of the theoretical models: MLR (pink), MKT (purple), and Ney (orange).}
    \label{fig:single_ya}   
\end{figure}

We first perform a baseline simulation with the mean half-lives predicted by the MDN in the unstable nuclei where experimental values are unavailable, comparing the resulting final abundance pattern with simulations based on theoretical half-lives (MLR, MKT, and Ney). As Fig.\ \ref{fig:single_ya} shows, the resulting abundance pattern reveals the impact of deviations in the predicted rates, particularly in the vicinity of the third $r$-process peak ($A \sim 195$).  The MDN shifts the peak toward lower-mass nuclei relative to the MLR and MKT baselines. The shift is similar to but smaller than that produced by the Ney rates, so that the MDN results are bracketed by those of the theoretical models.

This behavior is a direct consequence of the data trends identified in the previous section.  The MDN predicts systematically longer half-lives for lower-$Z$ nuclei at the $N=126$ shell closure than do the MKT and MLR models. Because material at this closed shell must undergo $\beta$-decay to continue capturing neutrons, the longer lifetimes act as a tight bottleneck. The $r$-process stalls, causing matter to accumulate at lower $Z$ along the isotonic chain. Since $A = Z + 126$, trapping material at lower $Z$ shifts the final abundance distribution, resulting in a third $r$-process peak with higher abundances occurring at smaller mass.

\begin{figure}
    \centering
    \includegraphics[width=\linewidth]{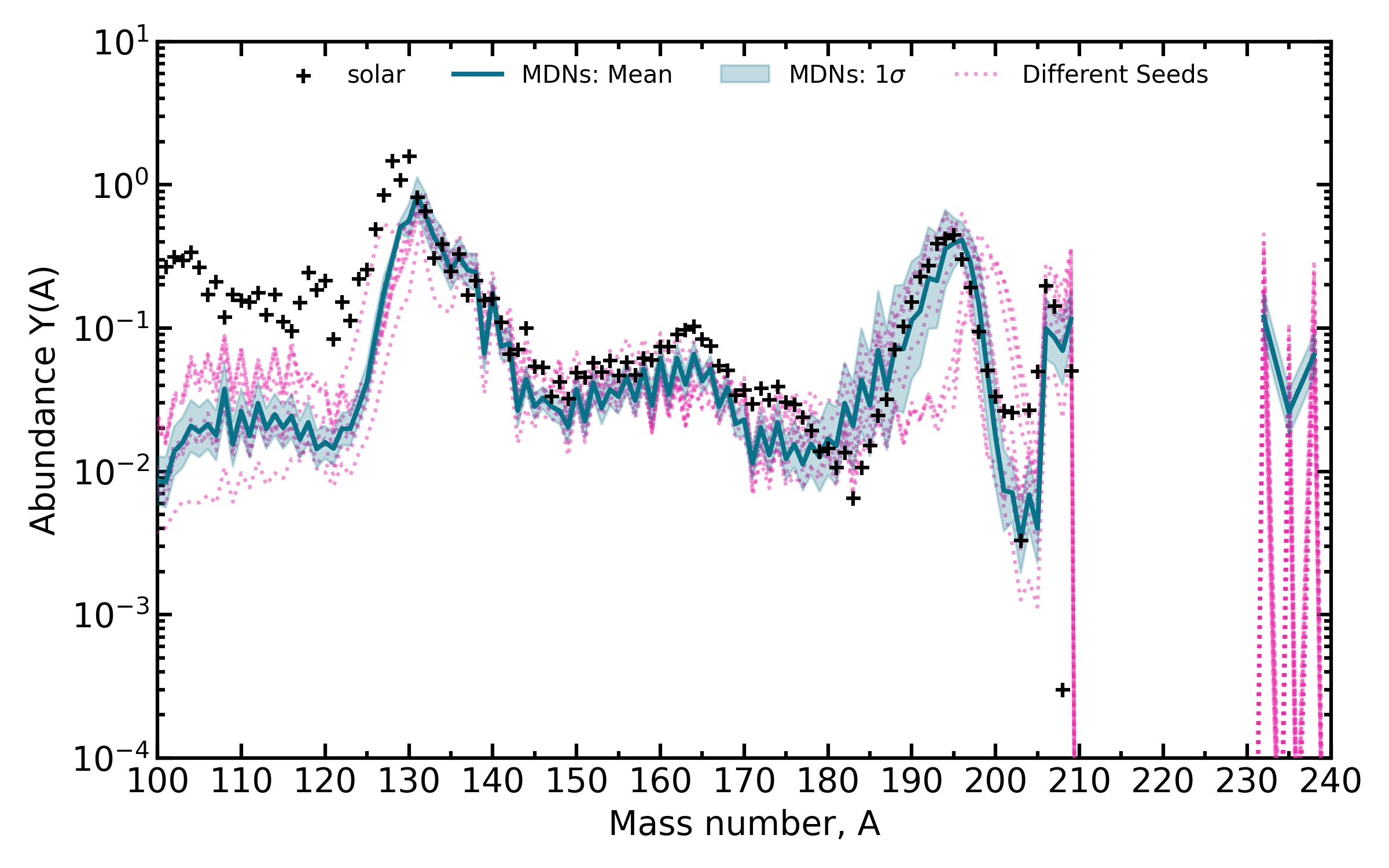}

    \includegraphics[width=\linewidth]{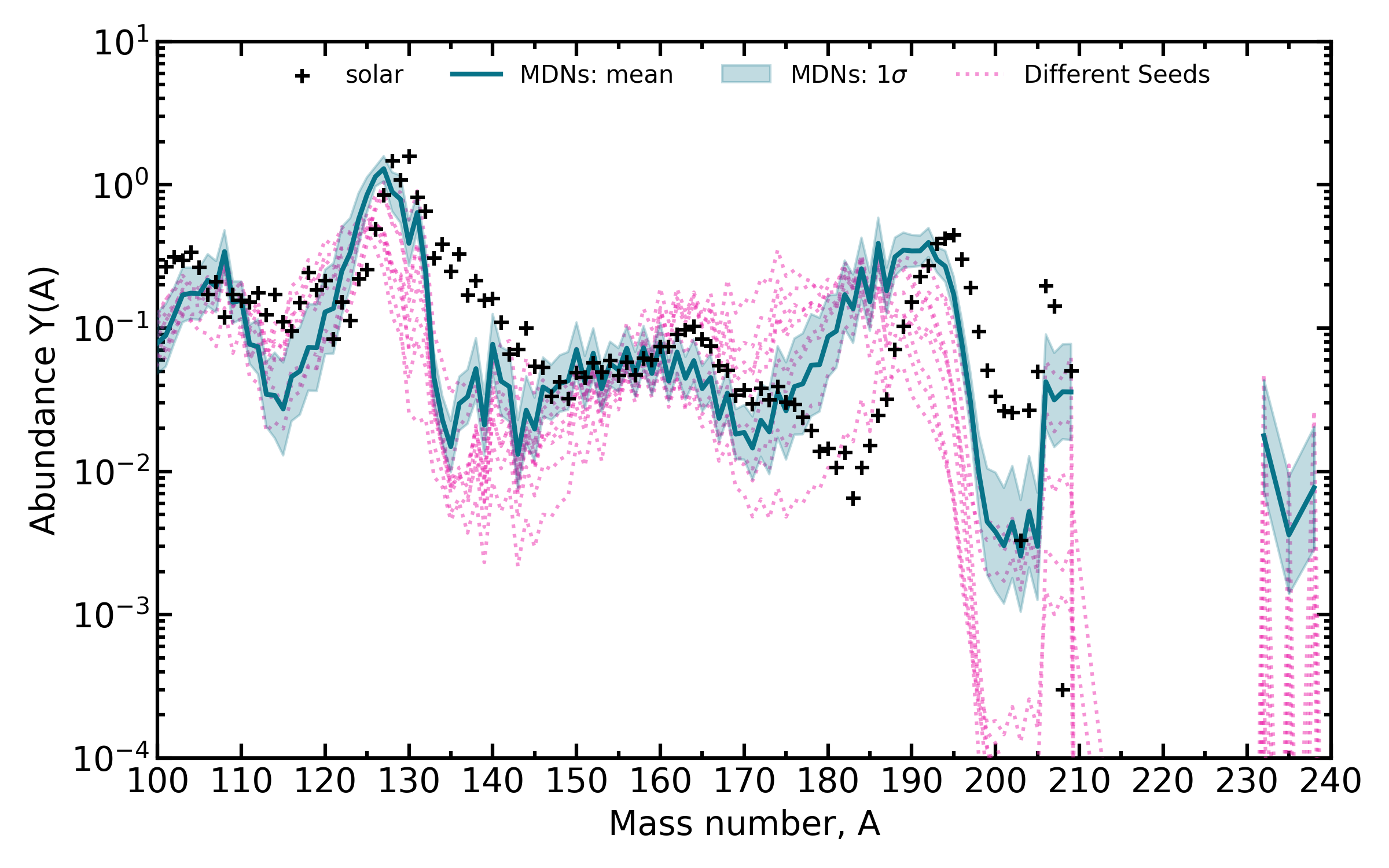}
    
    \caption{The simulated $r$-process abundance uncertainty arising from half-life uncertainties. 
    The dotted lines show the seed-to-seed variations.
    Top: Cold very neutron-rich ($Y_e = 0.02$) conditions. Bottom: Moderately neutron-rich conditions.}
    \label{fig:ya_band_combined}
\end{figure}

One advantage of the MDN approach is the ability to quantify predictive uncertainty, allowing us to propagate nuclear physics uncertainties into astrophysical observables.
To do so, we use Monte Carlo sampling to generate 100 independent sets of half-lives for the entire nuclear chart from a single choice of the trained MDN model. In each set, we choose the half-life of every isotope from its MDN probability distribution.  We then perform the $r$-process network simulation with each of these 100 unique global beta-decay sets. The resulting mean and $1\sigma$ uncertainty band for the final isotopic abundances appear in the top panel of Fig.\ \ref{fig:ya_band_combined}.  The variance (teal shade) represents the uncertainty in nucleosynthesis yields from the intrinsic error in our global beta-decay rates associated, as we have said, with one MDN model.  Even with that restriction, the variance is significant.
Importantly, the uncertainty is not the same everywhere, and is largest near the third abundance peak ($A \sim 195$). The increase in uncertainty at large mass reflects the scarcity of half-life data there. Future measurements in the regions where the uncertainty is largest would best constrain $r$-process models.

Because we are interested in other environments, we repeat the uncertainty analysis for a cold wind scenario, with viscously driven outflows that are typical of those from an accretion disk. This environment produces higher electron fractions, and we take $Y_e = 0.19$ and $Y_e = 0.23$ to account for the formation of lighter and heavier $r$-process elements, respectively. In part because of the higher $Y_{e}$, the abundance pattern, shown in Fig.\ \ref{fig:ya_band_combined}, differs significantly from that in Fig.\ \ref{fig:single_ya}. The comparison also shows that the impact of $\beta$-decay uncertainties is context-dependent; changing the thermodynamic history of the ejecta shifts the nucleosynthetic flow, with results that reflect the mass-dependence of the MDN variance.

Finally, for comparison, Fig.\ \ref{fig:ya_band_combined} also overlays the abundance patterns from models trained with different sets of data (shown in pink).  The spread in final $r$-process abundances caused by changing the training data (initialized by seed number) is larger than the intrinsic uncertainty derived from a single model. This large variance, which arises from the neural-network initialization and training, represents a form of extrapolation uncertainty distinct from that presented earlier.

\subsection{Kilonova Luminosity}
\label{subsec:light_curves}

Having seen the effects on the $r$ process of uncertainties in $\beta$-decay half-lives, we now propagate the predictions for half lives to another important astrophysical observable: the kilonova light curve. The observable luminosity of a kilonova is driven by the radioactive decay of the newly-formed $r$-process elements, particularly the lanthanides, which play a crucial role in powering the late-time ``red" component of the light curve because of their high opacities. To quantify their effects, we first calculate the total effective nuclear heating rate, $\dot{Q}(t)$, with Eq. \ref{eq:Q_dot}, by coupling the energies released in each decay channel with the thermalization efficiency of the expanding ejecta.

\subsubsection{Effective Nuclear Heating Rates}

\begin{figure}[h!]
    \centering
    \includegraphics[width=\linewidth]{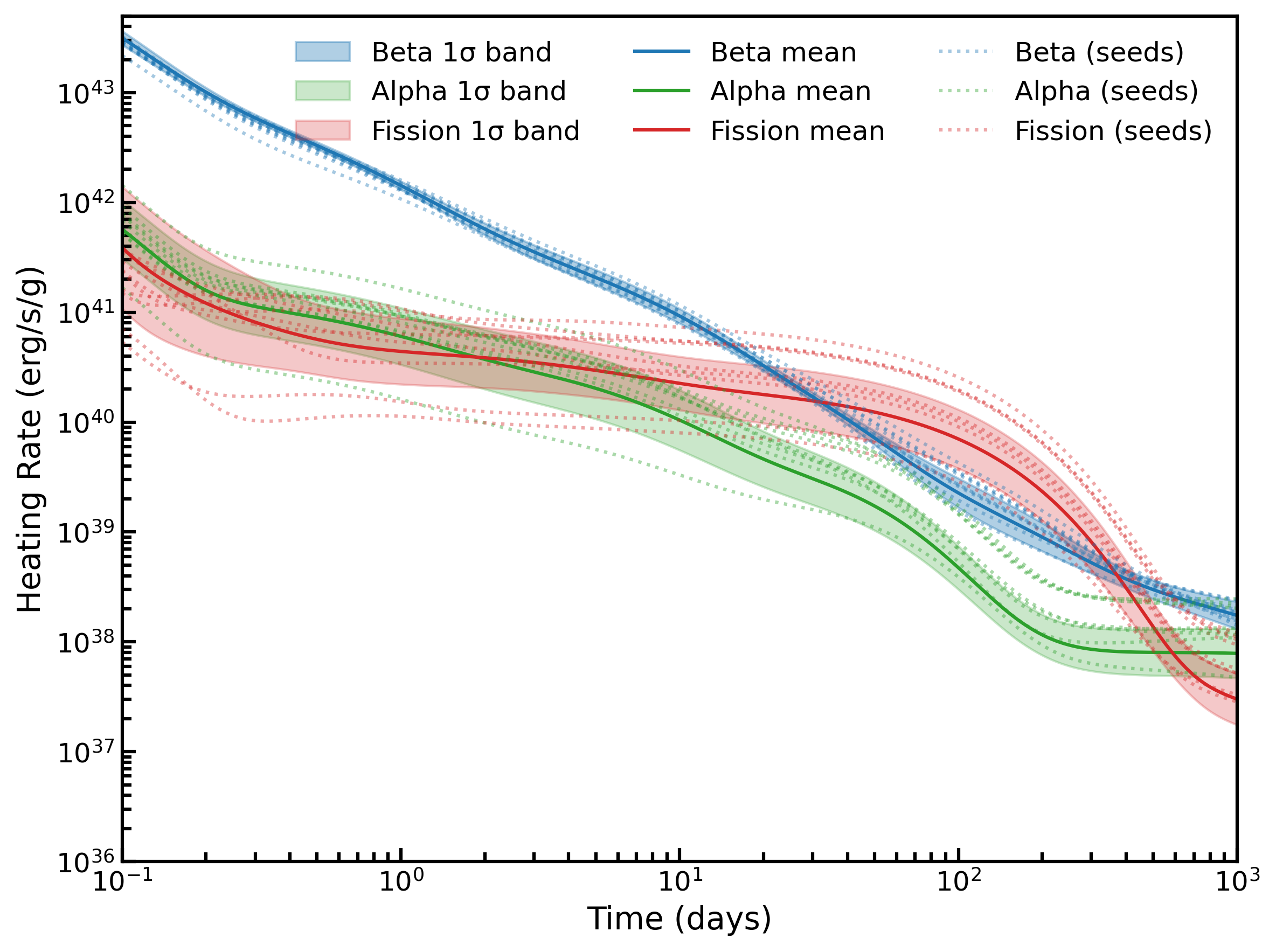}
    
    \includegraphics[width=\linewidth]{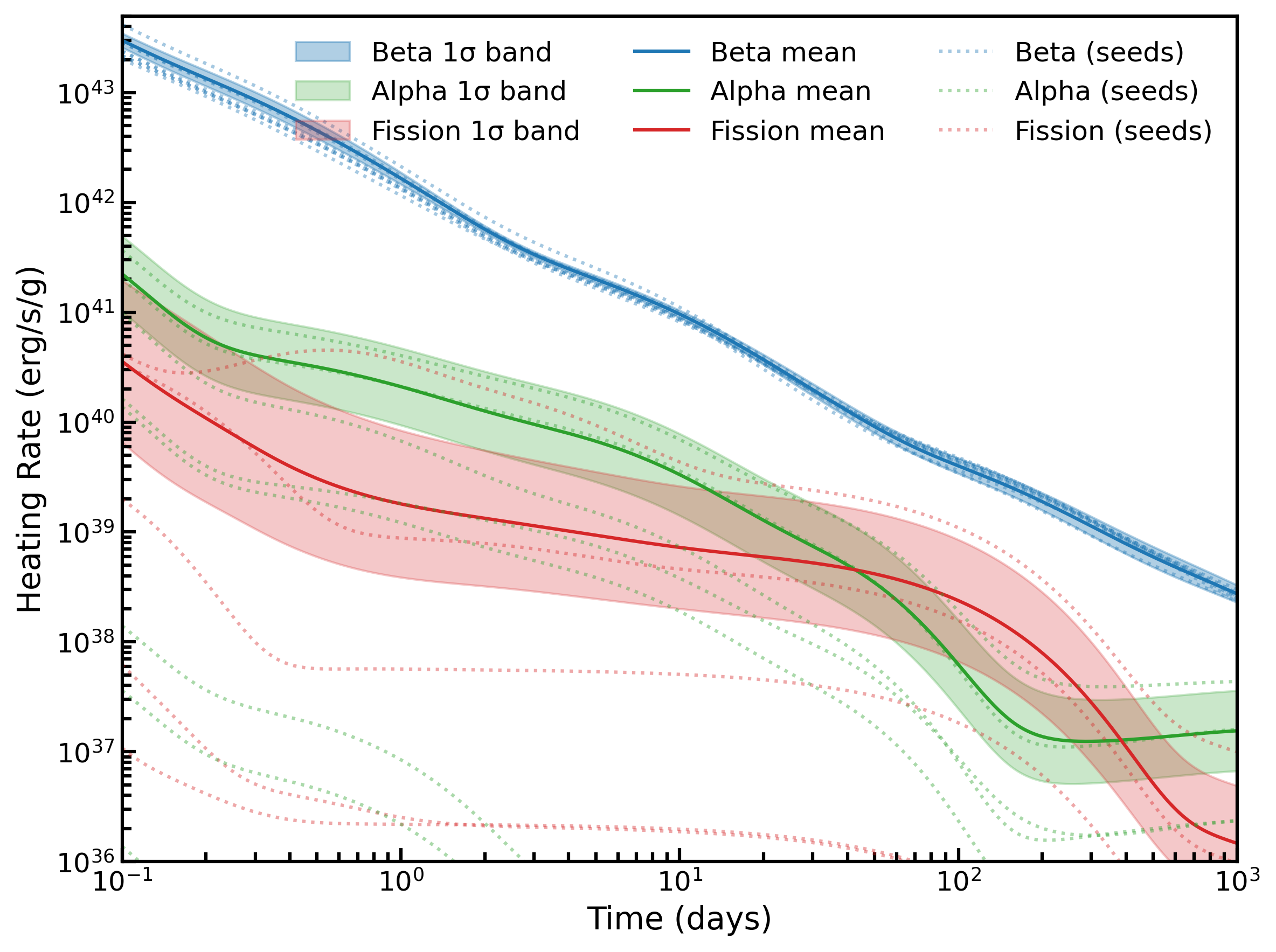}
    
    \caption{Evolution of the effective nuclear heating rates. The total heating is decomposed into contributions from $\beta$-decay (blue), $\alpha$-decay (green), and spontaneous fission (red). The spreads represent the uncertainty propagated from the MDN-predicted $\beta$-decay half-lives. The dotted lines show the seed-to-seed variations. Top: The cold low $Y_e$ scenario. Bottom: The cold moderate $Y_e$ scenario.}
    \label{fig:hr_comp_combined_filled}   
\end{figure}

By applying this thermalization formalism to the dynamical-ejecta scenario, we can isolate the energy sources driving the early and late-time light curves. Figure \ref{fig:hr_comp_combined_filled} breaks the total effective heating rate into components produced by $\beta$-decay (blue), $\alpha$-decay (green), and spontaneous fission (red). To capture the propagation of intrinsic nuclear uncertainties, we show the mean and $1\sigma$ uncertainty band of the heating rates, computed from an ensemble of $r$-process network simulations with $\beta$-decay rates sampled directly from the MDN probability distributions. For comparison, we also show the results of using different sets of training data for each channel in dotted lines.

The evolution of the heating rate reveals a clear hierarchy in the energy sources. At early times ($t < 10$ days), the heating is dominated by $\beta$-decay, which provides the primary power for the peak of the light curve. During this phase, contributions of $\alpha$-decay and fission contributions are small, with that of $\alpha$-decay being comparable to fission but declining rapidly.

A critical transition occurs at later times ($t > 10$ days), however. As the $\beta$-decay intensity continues to drop, the fission channel emerges as the dominant heating source. This late-time dominance has been identified previously as the signature of spontaneous fission from trans-lead nuclei, specifically $^{254}$Cf \citep{Zhu_2021}. The fission heating can last for hundreds of days.

We next examine the heating rates for the disk wind scenario, characterized by a moderate electron fraction. The results are in the bottom panel of Fig.\ \ref{fig:hr_comp_combined_filled}.  Like in the dynamical ejecta, $\beta$ decay is the dominant energy source at early times. The contributions of $\alpha$-decay and spontaneous fission are different, however.  Because the disk wind is only moderately neutron rich, the nucleosynthesis does not populate the actinides as strongly as in dynamical ejecta. 
The heating contribution from fission and $\alpha$-decay is thus markedly lower.

\subsubsection{Bolometric Light Curves}

The variance in our calculated heating rates ultimately shows up as uncertainty in the observable electromagnetic transient. Because the temporal evolution of the nuclear heating dictates the shape, peak luminosity, and late-time behavior of the kilonova \citep{Barnes_2021, Zhu_2021, Lund_2023}, we must propagate our $\beta$-decay uncertainties through to the final emission. Following the thermodynamic approach described in the Methods section \ref{sec:methods}, we compute the corresponding bolometric light curves over a 40-day post-merger time.

\begin{figure}[h!]
    \centering
    \includegraphics[width=\linewidth]{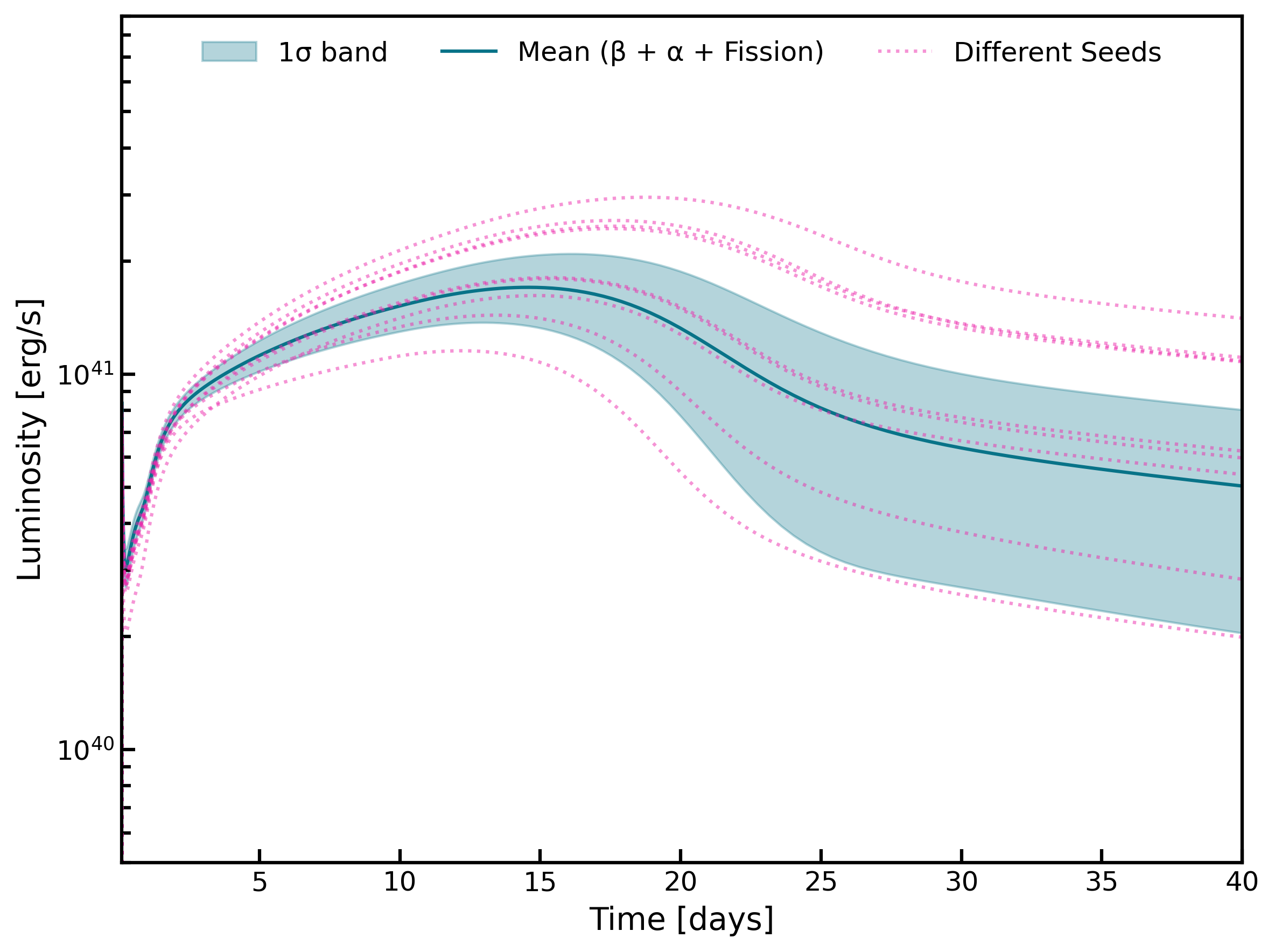}

    \includegraphics[width=\linewidth]{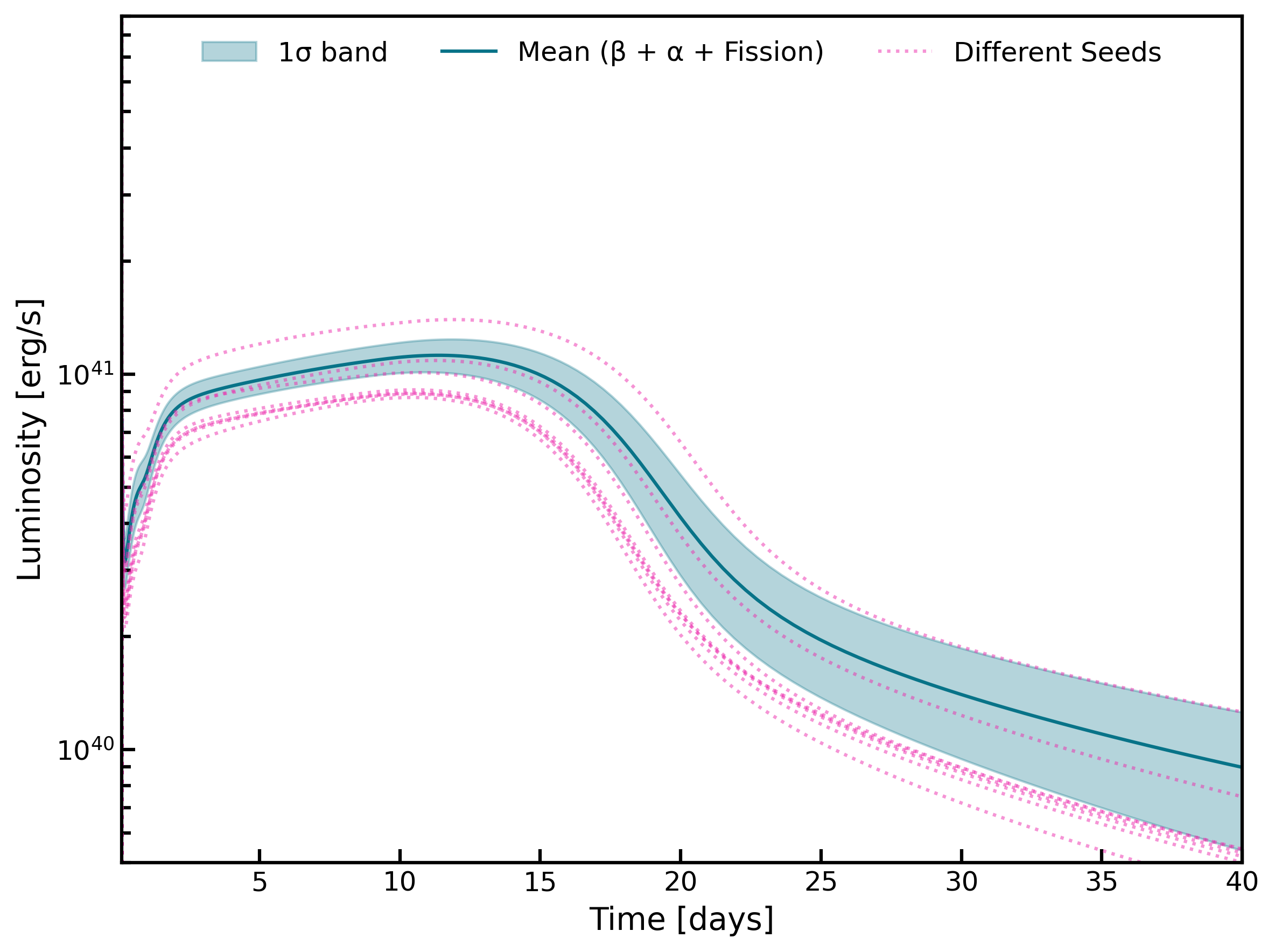}
    
    \caption{\ML{Simulated bolometric kilonova light curves. The shaded regions represent the 1 $\sigma$ uncertainty band, driven by the variance in the MDN $\beta$-decay rates. Top: Cold low $Y_e$ scenario. Bottom: Disk wind moderate $Y_e$ scenario. In both panels, the overlaid dotted pink lines represent predictions from models trained with different random seeds.}}
    \label{fig:lc_band_combined}   
\end{figure}

The simulated bolometric light curves for the dynamical ejecta appear in the top panel of Fig.\ \ref{fig:lc_band_combined}. 
The uncertainties in the light-curve evolution are driven by variations in the nuclear heating rates, which stem directly from the underlying variance in our MDN $\beta$-decay predictions. At early times ($t < 10$ days), the luminosity uncertainty remains relatively low, reflecting the tighter constraints on the dominant $\beta$-decay flow. At later times, however, the highly uncertain contribution of spontaneous fission becomes comparable to, and eventually dominates, that of $\beta$ decay. 
The change is responsible for two observational effects. 
First, the substantial injection of fission energy sustains a higher luminosity, producing a prolonged luminous tail rather than a rapid decline. 
Second, the variance in the light curve expands significantly, directly mirroring the larger uncertainties in late-time fission heating.

The bottom panel of Fig.\ \ref{fig:lc_band_combined} illustrates the predicted light curve evolution in the disk wind scenario. Because the nuclear heating is predominantly due to $\beta$-decay at all times, the propagated variance remains comparatively small compared with that in the dynamical case.  The overall luminosity band is thus more tightly constrained. Lacking the late-time energy injection from spontaneous fission that characterizes dynamical ejecta, this wind cannot sustain a prolonged luminous tail. Instead, the luminosity declines rapidly following its peak at approximately 10 days.

Even with the narrower intrinsic uncertainty band in moderately neutron-rich conditions, the predicted late-time luminosity still varies widely.  
Furthermore, as the overlaid curves on both panels show, changing the training data (initialization seed number) increases the uncertainty. 
Decay half-life extrapolations in data-poor regions can therefore significantly affect kilonova modeling, even in less extreme ejecta environments.
Ultimately, we will need new experimental measurements to constrain extrapolation models and better understand astrophysical observables.

\section{Discussion and Conclusion}

In this paper, we introduced a Mixture Density Network (MDN) framework to predict $\beta$-decay half-lives and quantify their intrinsic uncertainties. Unlike standard deterministic models, the MDN parameterizes the probability distribution of half-lives, providing a dynamically-scaled intrinsic (aleatoric) uncertainty. 
Our framework achieves high predictive accuracy on known nuclei, yielding an overall RMSE of 0.597 on the NUBASE2020 dataset, while theoretical models such as MLR, MKT, and Ney have RMSE values larger than 0.8.

We also considered the important challenge of extrapolating into data-poor regions far from stability. There, variations in ML training data can introduce significant epistemic (parameter) uncertainty, leading to divergent predictions, particularly in the critical $N=126$ isotonic chain. 
Recognizing that model extrapolations diverge greatly with different seed choices, we focused primarily on a single baseline model to evaluate the astrophysical impact of intrinsic MDN uncertainties, and discussed the effects of varying seeds only briefly.

Next, we carried out $r$-process network calculations to evaluate the impact of the MDN half-lives on heavy element nucleosynthesis. The inherent variance in $\beta$-decay rates turned out to have a significant effect on abundance patterns, particularly in and around the third $r$-process peak. 

We then examined the effects of the nuclear uncertainties on astrophysical observables. By coupling the isotopic abundances with the time-dependent thermalization efficiencies of decay products, we calculated the effective nuclear heating rates in homologously expanding ejecta (with $M_{\text{ej}} = 0.05 M_\odot$, $v_{\text{ej}} = 0.15c$).  Using a temperature-dependent opacity suitable for lanthanide-rich environments, we translated the heating rates into bolometric kilonova light curves, which showed how intrinsic uncertainties from a MDN $\beta^-$ decay half-life model shape the late-time transient.

Although our framework provides data-driven constraints on the $r$-process, it has limitations.
Because we explicitly isolated the intrinsic (aleatoric) variance of one $\beta$ half-life MDN model, we did not fully propagate the epistemic (parameter) uncertainty of the neural network. 
The total uncertainty bands in the most neutron-rich regimes are therefore probably underestimated. Ultimately, reducing the epistemic uncertainty requires more than purely theoretical masses and half-lives. 
Future experimental measurements of neutron-rich isotopes at facilities such as FRIB \citep{Schatz_2022}, ARIEL\citep{GARNSWORTHY_2019}, HIAF \citep{Yang_2013}, RIBF\citep{Phong_2022}, FAIR \citep{Caballero_2016} and the N=126 Factory \citep{Valverde_2026} will be essential for providing stringent constraints on the ML training data. 
Improved versions of our approach can complement experiment by incorporating physics-informed loss functions to naturally suppress unphysical parameter divergence during training, and provide more accurate predictions near the neutron drip line.

\section{Acknowledgments}
M. L., F. W. and R. S. acknowledge support from the Network for Neutrinos, Nuclear Astrophysics and Symmetries (N3AS), through the National Science Foundation Physics Frontier Center award No. PHY-2020275.
N. V. acknowledges the support of the Natural Sciences and Engineering Research Council of Canada (NSERC). TRIUMF receives federal funding via a contribution agreement with the National Research Council (NRC) of Canada.
R. S. additionally acknowledges support from the U.S. Department of Energy under Grant Nos. DE-FG02-95-ER40934, LA22-ML-DE-FOA-2440, and DE-SC00268442 (ENAF).
J. E. acknowledges support from the U.S.\ Department of Energy under Grant Nos.\ DE-FG02-97ER41019 and DE-SC0023495.
\bibliographystyle{aasjournalv7}
\bibliography{ref}

\end{document}